\documentclass[prl,nofootinbib,twocolumn,superscriptaddress]{revtex4}
\def\mysection#1{{\bf #1.} }
\def\mysections#1{{\bf #1.} }

\newcommand{\be}{\begin{equation}}
\newcommand{\ee}{\end{equation}}
\newcommand{\bea}{\begin{eqnarray}}
\newcommand{\eea}{\end{eqnarray}}
\newcommand{\beq}{\begin{equation}}
\newcommand{\eeq}{\end{equation}}
\newenvironment{Eqnarray}%
         {\arraycolsep 0.14em\begin{eqnarray}}{\end{eqnarray}}
\def\beqa{\begin{Eqnarray}}
\def\eeqa{\end{Eqnarray}}
\newcommand{\no}{\nonumber}
\def\lsim{\mathrel{\rlap{\lower4pt\hbox{\hskip1pt$\sim$}}
    \raise1pt\hbox{$<$}}}         
\def\gsim{\mathrel{\rlap{\lower4pt\hbox{\hskip1pt$\sim$}}
    \raise1pt\hbox{$>$}}}         
\begin{document}

{\hspace*{13cm}\vbox{\hbox{WIS/17/03-Jul-DPP}
  \hbox{hep-ph/0307081}}}

\vspace*{-10mm}

\title{\boldmath Leptogenesis from Supersymmetry Breaking}

\author{Yuval Grossman}\email{yuvalg@physics.technion.ac.il}
\affiliation{Department of Physics, Technion--Israel Institute of
  Technology, Technion City, 32000 Haifa, Israel}

\author{Tamar Kashti}\email{tamar.kashti@weizmann.ac.il}
\affiliation{Department of Particle Physics,
  Weizmann Institute of Science, Rehovot 76100, Israel}

\author{Yosef Nir}\email{yosef.nir@weizmann.ac.il}
\affiliation{Department of Particle Physics,
  Weizmann Institute of Science, Rehovot 76100, Israel}

\author{Esteban Roulet}\email{roulet@cab.cnea.gov.ar}
\affiliation{Centro At\'omico Bariloche,
  Av. Bustillo 9500, 8400 Bariloche, Argentina}

\vspace*{1cm}

\begin{abstract}
We show that soft supersymmetry breaking terms involving the heavy
sneutrinos can lead to sneutrino-antisneutrino mixing and to new
sources of CP violation, which are present even if a single generation
is considered. These terms are naturally present in supersymmetric
versions of leptogenesis scenarios, and they induce indirect CP
violation in the decays of the heavy sneutrinos, eventually generating
a baryon asymmetry. This new contribution can be comparable to or even
dominate over the asymmetry produced in traditional leptogenesis scenarios.
\end{abstract}

\maketitle

\mysection{Introduction}
Leptogenesis \cite{Fukugita:1986hr} provides an attractive scenario
for the generation of the baryon asymmetry of the Universe. The
three necessary conditions \cite{Sakharov:dj} are fulfilled in the 
Standard Model with additional, heavy, singlet neutrinos: the
Majorana masses for singlet neutrinos violate lepton number (which is
then reprocessed into baryon number by the sphaleron processes), their
Yukawa interactions with lepton doublets violate CP, and
their decay can occur out-of-equilibrium. The same framework can
account for the observed neutrino masses through the see-saw mechanism
(for a review, see \cite{Gonzalez-Garcia:2002dz}).  

This standard leptogenesis scenario works similarly in the
supersymmetric generalization of the Standard Model (SSM+N), where it
is based on superpotential terms  
\beq\label{suppot}
W=\frac12M_i N_iN_i+Y^\nu_{ij}L_iN_jH_u.
\eeq
Here $N_i$, $L_i$ and $H_u$ stand for, respectively, the singlet
neutrino, lepton doublet and up-Higgs doublet chiral superfields. The
masses $M_i$ provide lepton number violation and the Yukawa
matrix $Y^\nu$ provides CP violation. Soft supersymmetry breaking
terms provide additional sources of lepton number and CP violation.
Particularly interesting are bilinear and trilinear scalar couplings
involving the singlet sneutrino fields:
\beq\label{lsoft}
{\cal L}_{\rm soft}=B_{ij} \widetilde N_i\widetilde N_j+A^\nu_{ij}\widetilde
L_i\widetilde N_j H_u.
\eeq
The effects of these terms are usually ignored because they are
assumed to be highly suppressed by the small ratio between the
supersymmetry breaking scale (say, $\tilde m\sim10^3$ GeV) and the mass
scale of the singlet neutrinos ($M\sim10^{10}$ GeV). 

In this work we investigate the effects of the soft supersymmetry
breaking terms (\ref{lsoft}) on leptogenesis. These soft terms clearly
affect only the asymmetry generated in the heavy sneutrino decays and,
as shown below, lead to CP violation even in a single sneutrino
generation framework. On the qualitative level, the effects that we
consider are unique among all scenarios of baryogenesis and
leptogenesis that involve late decays of heavy particles: they come
from {\it indirect} CP violation, that is CP violation in mixing and
in the interference of decays with and without mixing,
similar to neutral meson decays. In contrast, previous scenarios and,
in particular, the standard leptogenesis, are based on direct CP
violation or on the mixing of (s)neutrinos of different
generations \cite{Fukugita:1986hr,Covi:1996wh,fl95,pi99}. 
On the quantitative level, we find surprising results:
there are regions in parameter space where the effects of the soft
supersymmetry breaking terms are non-negligible due to the enhancement
produced by sneutrino mixing effects and could
even be the dominant source of leptogenesis.

\mysection{Mixing and Decays of Singlet Sneutrinos}
Since we are interested in the effects of supersymmetry breaking, we
work in a simplified single generation model. In the three generation
case, the relevant out of equilibrium decays are usually those of the
lightest heavy neutrino and sneutrino, while the decays of the heavier
ones play no role. Our simplified single generation model would then
refer to the lightest of the three heavy sneutrinos.
The relevant terms in the Lagrangian are the following:
\beqa\label{lagrangian}
{\cal L}&=&|M|^2\widetilde N^\dagger\widetilde N+\left[M NN+B\widetilde N\widetilde
  N+{\rm h.c.}\right]\no\\ 
&+&\left[Y\widetilde N Lh+Y \widetilde L Nh+ Y H NL+{\rm h.c.}\right]\no\\
&+&\left[MY^{*}\widetilde N\widetilde L^\dagger H^\dagger+A\widetilde N\widetilde
  L H+{\rm h.c.}\right] .
\eeqa
Here $\widetilde N,\widetilde L,H$ are scalar fields, while $L,N,h$
are their fermionic superpartners. The first line includes mass terms,
the second Yukawa couplings, and the third trilinear scalar
couplings. For simplicity, we omit the
subscript from $H_u$ and the superscripts from $Y^\nu$ and $A^\nu$.

The Lagrangian of eq. (\ref{lagrangian}) has a single physical CP
violating phase:
\beq\label{cpvpha}
\phi_N=\arg(AMB^*Y^{*}).
\eeq
This phase would give the CP violation that is necessary to
dynamically generate a lepton asymmetry. If we set the lepton number of
$N$ and $\widetilde N$ to $-1$, so that $Y$ and $A$ are lepton
number conserving, the two couplings $M$ and $B$ violate lepton number
by two units. Thus processes that involve $Y$ or $A$, and $M$ or
$B$, would give the lepton number violation that is necessary for
leptogenesis.

A crucial role in our results is played by the $\widetilde N-\widetilde
N^\dagger$ mixing amplitude \cite{Grossman:1997is} ,
\beq\label{defmonetwo}
\langle \widetilde N|{\cal H}|\widetilde N^\dagger\rangle=M_{12}-\frac
i2\Gamma_{12},
\eeq
which induces a mass difference $\Delta M$ and a width
difference $\Delta\Gamma$ between the two mass eigenstates,
\beq\label{defpq}
|\widetilde N_{L,H}\rangle=p|\widetilde N\rangle\pm q|\widetilde
N^\dagger\rangle,\ \ \ \ R\equiv|q/p|.
\eeq
CP violation in mixing is signaled by $R\neq1$.

There are four relevant decay amplitudes:
\beqa\label{decamp}
A_{\widetilde L}&=&\langle \widetilde L H|{\cal H}|\widetilde N\rangle,\ \ \
\bar A_{\widetilde L^\dagger}=\langle \widetilde L^\dagger 
H^\dagger|{\cal H}|\widetilde N^\dagger\rangle,\no\\
\bar A_{L}&=&\langle Lh|{\cal H}|\widetilde N^\dagger\rangle,\ \ \
A_{\bar L}=\langle \bar L\bar h|{\cal H}|\widetilde N\rangle.
\eeqa
To simplify the discussion and emphasize the more
significant results of our analysis, we make several assumptions and
approximations. (The complete treatment, including smaller effects
that we neglect here, will be presented elsewhere \cite{future}.)
\begin{enumerate}
\item We assume that there is no initial asymmetry in the number
  densities of $\widetilde N$ and $\widetilde N^\dagger$. Such an asymmetry
  requires physics beyond the SSM+N \cite{Allahverdi:2003tu}.
  Thus, as long as thermal equilibrium persists, the number densities
  remain equal.
\item We neglect direct CP violation. The direct CP
  violation that drives the standard leptogenesis has been switched
  off by working in a single generation model. We have checked that
  all the effects of direct CP violation that remain in the framework
  of the Lagrangian (\ref{lagrangian}) can be safely neglected
  \cite{future}. This assumption implies that CP-conjugate amplitudes in
eq. (\ref{decamp}) have equal magnitudes:
\beq\label{decampe}
A_{\widetilde L}=\bar A_{\widetilde L^\dagger},\ \ \
A_{\bar L}=\bar A_L.
\eeq
\item We neglect the small decay amplitudes that arise only when
  supersymmetry is broken (except for their contribution to
  $\Gamma_{12}$). Thus we put
  \beq\label{zeramp}
\langle \widetilde L H|{\cal H}|\widetilde
N^\dagger\rangle=\langle \widetilde L^\dagger H^\dagger|{\cal
  H}|\widetilde N\rangle=0.
\eeq  
\end{enumerate}  

We simplify our treatment by assuming an instantaneous departure from
equilibrium at $t=0$. Defining $|\widetilde N(t)\rangle$ and
$|\widetilde N^\dagger(t)\rangle$ to be the states that evolve from
purely $|\widetilde N\rangle$ and $|\widetilde N^\dagger\rangle$
(assumed to be the initial states at $t=0$), we
obtain the following rates: 
\beqa\label{tdrates}
\Gamma(\widetilde N(t)\to\widetilde L H)&=&\Gamma(\widetilde N^\dagger(t)\to\widetilde L^\dagger
H^\dagger)=\widetilde{\cal N}_2|A_{\widetilde L}|^2 g_+(t),\no\\
\Gamma(\widetilde N^\dagger(t)\to\widetilde L H)&=&\widetilde{\cal
    N}_2|A_{\widetilde L}|^2 R^{-2} g_-(t),\no\\
\Gamma(\widetilde N(t)\to\widetilde L^\dagger H^\dagger)&=&\widetilde{\cal N}_2|
A_{\widetilde L}|^2 R^{+2} g_-(t),\no\\
\Gamma(\widetilde N(t)\to\bar L\bar h)&=&\Gamma(\widetilde N^\dagger(t)\to
Lh)={\cal N}_2|A_{\bar L}|^2g_+(t),\no\\ 
\Gamma(\widetilde N^\dagger(t)\to\bar L\bar h)&=&{\cal
    N}_2|A_{\bar L}|^2 R^{-2} g_-(t),\no\\
\Gamma(\widetilde N(t)\to Lh)&=&{\cal N}_2|A_{\bar L}|^2 R^{+2} g_-(t).
\eeqa
Here $\widetilde{\cal N}_2$ and ${\cal N}_2$ are normalization
factors that include, in particular, the two body phase space factors, and
\beq\label{defht}
g_\pm(t)\equiv e^{-\Gamma t}\left[
  \cosh\frac{\Delta\Gamma t}{2}\pm\cos(\Delta M\ t)\right].
\eeq

\mysection{Lepton Asymmetry}
The lepton asymmetry is defined to be the following ratio:
\beq\label{defal}
\varepsilon_\ell\equiv\frac{\Gamma(\widetilde N,\widetilde N^\dagger\to\widetilde
  L,L+X)-\Gamma(\widetilde N,\widetilde N^\dagger\to\widetilde L^\dagger,\bar
  L+X)}{\Gamma(\widetilde N,\widetilde N^\dagger\to\widetilde
  L,L+X)+\Gamma(\widetilde N,\widetilde N^\dagger\to\widetilde L^\dagger,\bar
  L+X)}.
\eeq
Obviously, both CP and lepton number have to be violated to give
$\varepsilon_\ell\neq0$.

Eq. (\ref{tdrates}) shows several cancellations among CP-conjugate
rates (these cancellations would be violated by the small supersymmetry
breaking contributions to the decay amplitudes that we neglected):
\beqa\label{cpcancel}
\Gamma(\widetilde N(t)\to\widetilde L H)-\Gamma(\widetilde
N^\dagger(t)\to\widetilde L^\dagger H^\dagger)&=&0,\no\\
\Gamma(\widetilde N(t)\to\bar L\bar h)-\Gamma(\widetilde N^\dagger(t)\to
Lh)&=&0.
\eeqa
The other pairs of CP-conjugate processes have the same
time-dependence:
\beqa\label{cpnocan}
\Gamma(\widetilde N^\dagger(t)\to\widetilde L H)&-&
\Gamma(\widetilde N(t)\to\widetilde L^\dagger H^\dagger)\no\\
&=&\widetilde{\cal N}_2|A_{\widetilde L}|^2(R^{-2}- R^{+2}) g_-(t),\no\\
\Gamma(\widetilde N(t)\to Lh)&-&\Gamma(\widetilde N^\dagger(t)\to\bar L\bar
h)\no\\
&=&{\cal N}_2|A_{\bar L}|^2(R^{+2}-R^{-2}) g_-(t).
\eeqa
It is clear then that the lepton asymmetry that would be generated in
our framework has CP violation in mixing, $R\neq1$, as its source.

The time-integrated lepton asymmetry (in the approximation $R\simeq1$)
is proportional to  
\beqa\label{timint}
\chi&\equiv& \frac{\int_0^\infty dt\ g_-(t)}{\int_0^\infty dt\
  [g_+(t)+g_-(t)]}=\frac{x^2+y^2}{2(1+x^2)}\ ;\no\\
x&\equiv&\frac{\Delta M}{\Gamma},\ \ \ 
y\equiv\frac{\Delta\Gamma}{2\Gamma}. 
\eeqa
Note that the mass- and width-splitting between the two
sneutrino mass eigenstates are purely supersymmetry
breaking effects: in the supersymmetric limit, the mass and the width
should be equal to that of the neutrino $N$.

A crucial point in evaluating the lepton asymmetry induced by singlet
sneutrino decays is that in the supersymmetric limit
$\widetilde{\cal N}_2|A_{\widetilde L}|^2={\cal N}_2|A_{\bar L}|^2$. Indeed,
this result is not surprising: since no lepton asymmetry is generated
here in neutrino decays, we expect that in the supersymmetric limit
no such asymmetry would be generated also in sneutrino decays. We
checked this result explicitly also for three body sneutrino
decays.\footnote{We disagree on this point with
  ref. \cite{Allahverdi:2003tu}, which did not take into account the Higgs and higgsino 
mediated three body decays.}

From eq. (\ref{cpnocan}) we learn then that to have a contribution
within our framework to the lepton asymmetry, we must have a
supersymmetry breaking effect that gives a difference between the
sneutrino decay rates into final leptons and sleptons,
$\widetilde{\cal N}_2|A_{\widetilde L}|^2\neq{\cal N}_2|A_{\bar L}|^2$.
Such an effect can be parametrized as follows:
      \beq\label{epsll}
      \delta_{\widetilde L\bar L}\equiv\frac{\widetilde{\cal N}_2|A_{\widetilde
          L}|^2-{\cal N}_2|A_{\bar 
          L}|^2}{\widetilde{\cal N}_2|A_{\widetilde L}|^2+{\cal
          N}_2|A_{\bar L}|^2}. 
      \eeq
Our final result for the lepton asymmetry is the following:
\beq\label{lepasy}
\varepsilon_\ell=\frac12\;\delta_{\widetilde L\bar
    L}\left(R^{+2}-R^{-2}\right)\chi.
\eeq
CP violation is encoded in the $(R^{+2}-R^{-2})$ factor, and lepton
number violation is encoded in the
factor $\chi$. Each of these two factors and, as argued above, also $\delta_{\widetilde
  L\bar L}$, require supersymmetry breaking. 

\mysection{Evaluating the Size of the Lepton Asymmetry}
Eq. (\ref{lepasy}) implies that, as expected, the mechanism that we
consider for generating the lepton asymmetry is suppressed by the
ratio between the supersymmetry breaking scale ($A$ or $B/M$) and the
scale of the singlet mass ($M$). Let us write down this dependence 
explicitly.

The width of the singlet sneutrino is given, to a good approximation,
by $\Gamma=|M||Y|^2/(4\pi)$.
For the mixing parameters in (\ref{lepasy}) we obtain
\beq\label{difference}
R^{+2}-R^{-2}={\cal I}m\left(\frac{AMY^*}{2 \pi B}\right), \ \ \
x=\frac{8\pi|B|}{|M|^2 |Y|^2},
\eeq
and $y\lsim x$. Assuming that $B\gsim AM$ (see, {\it e.g.}
\cite{Kaplunovsky:1993rd}) and $\sin\phi_N={\cal O}(1)$,
we obtain an order of magnitude estimate for the CP violating and
lepton number violating factors in (\ref{lepasy}) which, for $x\lsim 
1$, becomes:
\beq\label{cpvlv}
\left(R^{+2}-R^{-2}\right)\left(x^2+y^2\right)\sim
\frac{32\pi AB}{M^3Y^3}.
\eeq
Notice that, although the asymmetry is quadratic in the soft breaking 
terms, the Yukawa couplings in the denominator will give a large
enhancement factor if they are small. 

As concerns the non-cancellation between final leptons and sleptons,
the effect must be supersymmetry breaking. At zero temperature, the
leading effect within the SSM+N is a shift of the sneutrino
mass-squared, which is of order $\tilde m^2/M^2$. At finite
temperature, however, or, more precisely, when $T\sim M$, there are
much larger contributions to $\delta_{\widetilde L\bar L}$, as discussed below.
 
\mysection{Finite Temperature Effects}
The smallness of $\delta_{\widetilde L\bar L}$ is related to the
cancellation between the sneutrino decay rates into final states of
opposite lepton numbers ({\it e.g.} $\widetilde N\to\widetilde LH$ and $\widetilde
N\to\bar L\bar h$). The effect of this cancellation is that a
sneutrino asymmetry (generated by some mechanism beyond the SSM+N
\cite{Allahverdi:2003tu} or by sneutrino mixing processes) is erased
by the decays. 

There is, however, an important ingredient that prevents this
cancellation, related to supersymmetry breaking by finite
temperature corrections \cite{Covi:1997dr}. Indeed, in the
thermal bath where the heavy sneutrinos decay (corresponding to 
typical temperatures $T/M\sim0.1-10$), bosonic and fermionic particles
have different occupation numbers due to their different
statistics. Analyzing the various thermal corrections, we find that
the dominant one arises from the final state factors corresponding to
Pauli blocking of final state fermions and Bose-Einstein stimulation
of decays into scalars. The decay $\widetilde N\to\bar L\bar h$ is suppressed by a
factor $(1-n_F)^2$, while the decay $\widetilde N\to\widetilde LH$ is enhanced
by a factor $(1+n_B)^2$, where
\beq\label{defnfb}
n_{F,B}=\frac{1}{e^{M/(2T)}\pm1}.
\eeq
For simplicity, we are considering the decaying sneutrinos at
rest. Incorporating these factors into the decay rates, we obtain
(neglecting sub-leading modes) the following relation:
\beqa\label{brfite}
{\rm BR}(\widetilde N\to\widetilde LH)&=&1-{\rm BR}(\widetilde N\to\bar L\bar
h)\no\\ &=&\frac{(1+n_B)^2}{(1-n_F)^2+(1+n_B)^2}.
\eeqa
We thus obtain
\beq\label{delll}
\delta_{\widetilde L\bar L}
\simeq\frac{(1+n_B)^2-(1-n_F)^2}{(1-n_F)^2+(1+n_B)^2}. 
\eeq
For $T/M=0.1,0.3,0.5,1$, this factor assumes numerical values of
$\delta_{\widetilde L\bar L}=0.013,0.36,0.64,0.88$, respectively. This
shows the importance of this effect, which completely lifts the
cancellation for temperatures in the range that is relevant for the decay.

In addition, at $T\sim M$ there exist ${\cal O}(1)$ thermal corrections
to $M_{12}$ and $\Gamma_{12}$ and ${\cal O}(g^2,Y^2)$ corrections
from the thermal masses acquired by the particles. We do not present
these corrections explicitly here because they do not introduce
significant changes to the overall picture. 

\mysection{Discussion}
The generated baryon to entropy ratio is given by
\beq\label{nbtos}
n_B/s\simeq -\kappa10^{-3}\varepsilon_\ell,
\eeq
where $\kappa\lsim1$ is a dilution factor which takes into account the
possible inefficiency in the production of the heavy sneutrinos or
erasure of the generated asymmetry by L-violating scattering
processes. (A more precise relation between the sneutrino parameters
and the $n_B/s$ ratio requires a solution of the 
Boltzmann equations \cite{pl98}.)
Since observations determine $n_B/s\sim10^{-10}$, we should require that
our mechanism yields $|\varepsilon_\ell|\sim10^{-6}$.
The question is whether
eq. (\ref{lepasy}) can naturally yield an asymmetry of this size.

The attractiveness of leptogenesis comes, in part, because the same
singlet neutrino parameters that fit, through the see-saw mechanism,
the light neutrino masses, are also able to account for the observed
baryon asymmetry (see {\it e.g.}
\cite{Buchmuller:2000as,Davidson:2003cq}). Explicitly, the condition
for the sneutrino to decay out of equilibrium (so that the parameter
$\kappa$ is not suppressed \cite{Fukugita:1986hr}) reads
$\Gamma<H(T=M)$, where $H\simeq1.66\sqrt{g_*}T^2/M_{\rm Pl}$ is the
Hubble rate. This condition constrains the light neutrino mass
parameter, $m\equiv Y^2v^2/M<2\times 10^{-3}$~eV ($v=246$ GeV) which,
in view of the present indications from solar and atmospheric
neutrinos, is an interesting range. 

Let us gain some further insight about the size of
$\varepsilon_\ell$ by considering generic features of string theories
\cite{Kaplunovsky:1993rd}. The most conservative assumptions
concerning $A$ and $B$ 
would be that $A={\cal O}(\tilde m Y)$ and $B={\cal O}(\tilde m
M)$. The scale $\tilde m$ is here the scale that characterizes the
soft supersymmetry breaking terms, say $\tilde m\sim10^3$ GeV. 
Yet, $A$ and $B$ may be different depending on the mechanism that
breaks supersymmetry (see for example, \cite{Hall:1997ah}).

The maximum value of the asymmetry,
\beq\label{maxasy}
\varepsilon^{max}_\ell\simeq\delta_{\widetilde L\bar L}\frac{A}{YM},
\eeq
is obtained for $x\simeq 1$, that is,
\beq\label{xeqone}
\frac{B}{M}\simeq \frac{\Gamma}{2}\lsim 2\ {\rm GeV}\left(\frac
{M}{10^9\ {\rm GeV}}\right)^2,
\eeq
where the last inequality results from
$\Gamma<H(T=M)$. Eq. (\ref{maxasy}) implies that, to generate 
$\varepsilon_\ell\gsim10^{-6}$, one needs $A/Y> 10^{-6}M$. For
$A/Y\simeq \tilde m\sim$~TeV, we obtain $M<10^9$~GeV. Eq.
(\ref{xeqone}) indicates then the need to consider small ($\ll\tilde
m$) values of $B/M$ \cite{D'Ambrosio:2003wy}.


We conclude that effects of supersymmetry breaking terms on
leptogenesis are not negligible.  For plausible values of the heavy
sneutrino mass and Yukawa couplings they  might even provide the
dominant source for the observed baryon asymmetry of the  universe.

{\bf Note added:} After this paper was submitted, a related paper has appeared
\cite{D'Ambrosio:2003wy}. To find the regions of parameter space where the
scenario can work, the authors of \cite{D'Ambrosio:2003wy} solved the Boltzmann
equations, concluding that small values of $B/M$ must be
considered. This work helped us to improve the final discussion of the results.

\mysections{Acknowledgments}
We thank Gian Giudice, Sacha Davidson, Alejandro Ibarra and Yael
Shadmi for useful discussions. This work was supported in part by a
grant from Fundaci\'on Antorchas/Weizmann.
The research of Y.G. and Y.N. is supported by a Grant from the G.I.F.,
the German--Israeli Foundation for Scientific Research and
Development, and by the Einstein Minerva Center for Theoretical Physics.
The work of Y.G. is supported by the United
States--Israel Binational Science Foundation through Grant
No.~2000133 and by the Israel Science Foundation under
Grant No.~237/01.
Y.N.\ is supported by the Israel Science Foundation founded by the Israel
Academy of Sciences and Humanities and by EEC RTN contract HPRN-CT-00292-2002.


\end{document}